\documentclass{aa}  

\makeatletter
\renewcommand*\aa@pageof{, page \thepage{} of \pageref*{LastPage}}
\makeatother
\newcommand{\dmunit}{pc\,cm$^{-3}$}
\newcommand{\Tasc}{T_{\rm asc}}
\usepackage{graphicx}
\usepackage{xcolor}
\usepackage{txfonts}
\usepackage{hyperref}
\begin{document}

   \title{Evidence of intra-binary shock emission from\\ the redback pulsar PSR\,J1048+2339\thanks{The results reported in this paper are based on observations carried out with ESO-VLT (0104.D-0589), SRT (17-19 and 52-19 proposals), LOFAR (id. 405751), Galileo, and {\it Swift} telescopes.}}


\author{A. Miraval Zanon\inst{1,}\inst{2}, P. D'Avanzo\inst{2}, A. Ridolfi\inst{3,}\inst{4}, F. Coti Zelati\inst{5,}\inst{6,}\inst{2}, S. Campana\inst{2}, C.\,Tiburzi\inst{7},  D. de Martino\inst{8}, T.\,Mu$\tilde{\rm n}$oz Darias\inst{9},  C.G. Bassa\inst{7}, L.\,Zampieri\inst{10}, A. Possenti\inst{3,}\inst{11}, F. Ambrosino\inst{12,}\inst{13,}\inst{14}, A.\,Papitto\inst{12}, M. C. Baglio\inst{15,}\inst{2},\\ M. Burgay\inst{3}, A. Burtovoi\inst{16,}\inst{10}, D.\,Michilli\inst{17,}\inst{7,}\inst{18,}\inst{19}, P.\,Ochner\inst{20,}\inst{10}, P. Zucca\inst{7}}

\institute{Universit\`a dell'Insubria, Dipartimento di Scienza e Alta Tecnologia, Via Valleggio 11, I-22100, Como, Italy
  \and INAF, Osservatorio Astronomico di Brera, Via E. Bianchi 46, I-23807, Merate (LC), Italy\\
  e-mail: \texttt{arianna.miraval@inaf.it}
    \and INAF, Osservatorio Astronomico di Cagliari, Via della Scienza 5, I-09047, Selargius (CA), Italy
  \and Max-Planck-Institut f$\ddot{\rm u}$r Radioastronomie, Auf dem H$\ddot{\rm u}$gel 69, D-53121, Bonn, Germany
  \and Institute of Space Sciences (ICE, CSIC), Campus UAB, Carrer de Can Magrans s/n, 08193, Barcelona, Spain
  \and Institut d'Estudis Espacials de Catalunya (IEEC), Carrer Gran Capit\`a 2--4, 08034, Barcelona, Spain
  \and ASTRON, The Netherlands Institute for Radio Astronomy, Oude Hoogeveensedijk 4, 7991 PD, Dwingeloo, the Netherlands
   \and INAF, Osservatorio Astronomico di Capodimonte, Salita Moiariello 16, I-80131, Napoli, Italy
  \and Instituto de Astrofisica de Canarias, Via Lactea, La Laguna E-38200, Santa Cruz de Tenerife, Spain
  \and INAF, Osservatorio Astronomico di Padova, Vicolo dell’Osservatorio 5, I-35122, Padova, Italy
  \and Universit\`a degli Studi di Cagliari, Dipartimento di Fisica, S.P. Monserrato-Sestu km 0,700, I-09042 Monserrato (CA), Italy
   \and INAF, Osservatorio Astronomico di Roma, Via Frascati 33, I-00078, Monteporzio Catone (Roma), Italy
  \and INAF, Istituto di Astrofisica e Planetologia Spaziali, Via Fosso del Cavaliere 100, I-00133, Roma, Italy
  \and Sapienza Universit\`a di Roma, Piazzale Aldo Moro 5, I-00185, Roma, Italy
  \and Center for Astro, Particle and Planetary Physics, New York University Abu Dhabi, PO Box 129188, Abu Dhabi, UAE
  \and Center of Studies and Activities for Space (CISAS) `G. Colombo', University of Padova, Via Venezia 15, I-35131, Padova, Italy
  \and Anton Pannekoek Institute for Astronomy, University of Amsterdam, Science Park 904, 1098 XH Amsterdam, The Netherlands
  \and Department of Physics, McGill University, 3600 University Street, Montr\'eal, QC H3A 2T8, Canada
  \and McGill Space Institute, McGill University, 3550 University Street, Montr\'eal, QC H3A 2A7, Canada
  \and Universit\`a degli Studi di Padova, Dipartimento di Fisica e Astronomia, Via F. Marzolo 8, I-35131, Padova, Italy
}

\authorrunning{A. Miraval Zanon et al.}
             
\bibliographystyle{aa}
   \date{Received; accepted}

 
\abstract{
 We present simultaneous multiwavelength observations of the 4.66 ms redback pulsar PSR\,J1048+2339. We performed phase-resolved spectroscopy with the Very Large Telescope (VLT) searching for signatures of a residual accretion disk or intra-binary shock emission, constraining the companion radial velocity semi-amplitude ($K_2$), and estimating the neutron star mass ($M_{\rm NS}$). Using the FORS2-VLT intermediate-resolution spectra, we measured a companion velocity of $291 < K_2 < 348$\,km\,s$^{-1}$ and a binary mass ratio of $0.209 < q < 0.250$. 
 Combining our results for $K_2$ and $q$, we constrained the mass of the neutron star and the companion to $(1.0 < M_{\rm NS} < 1.6)\,{\rm sin}^{-3}i\,M_{\odot}$ and $(0.24 < M_2 < 0.33)\,{\rm sin}^{-3}i\,M_{\odot}$, respectively, where $i$ is the system inclination. 
 The Doppler map of the H$\alpha$ emission line exhibits a spot feature at the
expected position of the companion star and an extended bright spot close to the inner Lagrangian point. We interpret this extended emission as the effect of an intra-binary shock originating from the
interaction between the pulsar relativistic wind and the matter leaving the companion star. The mass loss from the secondary star could be either due to Roche-lobe overflow or to the ablation of its outer layer by the energetic pulsar wind. Contrastingly, we find no evidence for an accretion disk. We report on the results of the Sardinia Radio Telescope (SRT) and  the Low-Frequency Array (LOFAR) telescope simultaneous radio observations at three different frequencies (150\,MHz, 336\,MHz, and 1400\,MHz). No pulsed radio signal is found in our search.  This is probably due to both scintillation and the presence of material expelled from the system which can cause the absorption of the radio signal at low frequencies. The confirmation of this hypothesis is given by another SRT observation (L-band) taken in 2019, in which a pulsed signal is detected. 
Finally, we report on an attempt to search for optical pulsations using IFI+Iqueye mounted at the 1.2\,m Galileo telescope at the Asiago Observatory.
}
   \keywords{pulsar: individual (PSR\,J1048+2339) -- stars: neutron -- X-ray: binaries -- binaries: spectroscopic
               }

  \maketitle
%

\section{Introduction}
Binary millisecond pulsars (MSPs) are believed to be the end product of the accretion process in low-mass X-ray binary systems (LMXBs) \citep{Alpar, Radhakrishnan}. MSPs attain short spin periods ($P_{\rm s} \lesssim 10$\,ms) by accreting matter and angular momentum from a companion star in a process called ``recycling'', which can last up to a few gigayears \citep{Tauris2013}. During this prolonged accretion phase, LMXB systems are observed as bright X-ray sources \citep{Campana1996}, mainly due to the presence of an accretion disk around the neutron star (NS). The optical spectra are characterized by strong, double-peaked emission lines of H and He produced in the optically thick accretion disk \citep{vanParadijs1995}. When the mass transfer rate decreases such a X-ray bright phase ends, double-peaked emission lines disappear, and emission in the radio and $\gamma$-ray bands, powered by the NS’s
rotation, becomes predominant. 
The recycling model has been strongly supported by observations of X-ray millisecond pulsations in transient accreting millisecond X-ray pulsars (AMXPs; \citealt{Wijnands}) and coherent X-ray pulsations during Type I X-ray bursts \citep{Strohmayer}. The ultimate proof of this evolutionary scenario came with the discovery of transitional millisecond pulsars (tMSPs, see \citealt{Papitto_deMartino} for a review), which are systems that have shown transitions between LMXB and radio MSP states on very short timescales \citep{Archibald, Papitto2013, Bassa2014}.

``Spider'' pulsars are a subclass of binary MSPs in which the pulsar is in a very tight orbit ($\lesssim 1$\,day) around a low-mass companion star that is losing mass \citep{Roberts2013}. The ionized gas lost by the companion can intercept and thus distort or even totally absorb the radiation coming from the NS, causing regular eclipses of the radio pulsar signal. After the discovery of the first eclipsing binary pulsar, PSR B1957+20 \citep{Fruchter1988_1, Fruchter1988}, many similar systems have been found both in globular clusters\footnote{\url{https://www3.mpifr-bonn.mpg.de/staff/pfreire/GCpsr.html}} and in the Galactic field\footnote{\url{https://apatruno.wordpress.com/about/millisecond-pulsar-catalogue/}}. Spider pulsars are divided into two further subpopulations \citep{Roberts2013}:  black widows (BWs) and redbacks (RBs). BW pulsars are characterized by a very low-mass ($\leqslant 0.1\,M_{\odot}$), semi-degenerate companion star. The companion mass loss is typically driven by the evaporation of the outer layer of the star, caused by the irradiation of the surface by the strong pulsar wind. RB systems have more massive nondegenerate companions (0.1-0.4\,$M_{\odot}$) that are only partially ablated.
Most systems of both classes show regular eclipses of their radio signal, which cover a small fraction (<20\%) of the orbit in the BWs and a more significant fraction (up to 80\%) of the orbit in the RBs \citep{Broderick2016, Polzin2020}.

These spiders are key systems to study the evolution of binary MSPs and even more to investigate the formation of isolated MSPs. All the tMSPs discovered so far, which represent an important breakthrough in the knowledge
of binary MSPs, are RB systems. Isolated MSPs might be originated from BW systems when their original companion star
disappears, getting completely ablated by energetic particles produced by the pulsar wind or by $\gamma$-ray particles powered by the pulsar spin-down energy \citep{Levinson, Tavani1993}. However, simulations indicate that the direct evaporation alone cannot be responsible for
the formation of most isolated MSPs \citep{Ginzburg}.
\cite{Ginzburg2020b} propose that the ablation wind when coupled to the companion magnetic field is able to remove orbital angular momentum from the binary through magnetic braking, maintaining a stable Roche-lobe overflow and thus leading to the total evaporation of the companion star.
These systems are also excellent laboratories to study a matter-radiation
interaction in the low-density and high magnetic field regime. These conditions are typical of the shocks that originated in the interaction between matter released by the companion star and the relativistic pulsar wind. Moreover, RBs and BWs are crucial to constrain the maximum mass of a NS (see \citealt{Linares2019} and \citealt{Strader} for a review). 

In this paper we investigate both these aspects, constraining the shock geometry and the NS mass in a RB system. The Doppler tomography method \citep{Marsh1988} allows us to derive two-dimensional velocity maps of the H$\alpha$ emission region, by combining spectra taken at different orbital phases along the whole orbit. This technique enables us to distinguish between an accretion disk pattern and the more complicated geometry and kinematics that are expected in the pulsar shock emission scenario. This method has not yet been exploited for RB systems, because only a few of them clearly show emission lines in their optical spectra during the radio pulsar state (for example PSR J1628--3205, \citealt{Cho}).

PSR J1048+2339 (hereafter J1048) was discovered with the Arecibo telescope in a search for high-latitude Fermi unidentified sources \citep{Cromartie2016}. \cite{Deneva} classified it as a RB system using multiwavelength observations. J1048 is a 4.66-ms-spinning NS with a spin-down power of 1.2$\times 10^{34}$\,erg s$^{-1}$. It is part of a tight binary system with an orbital period
of 6.01 hours and it is located at a distance of about 1.7 kpc \citep{Deneva2020}. Its companion star has a mass of $\sim$0.4\,$M_{\odot}$ and an effective temperature $T_{eff} \sim$ 4200\,K \citep{Yap}. 
The inclination of the system is not well-constrained. \cite{Yap} fix the inclination to a theoretical upper limit of 76$^{\circ}$ because X-ray eclipses in \textit{Chandra} data are not observed.
Contrastingly, spectroscopic studies from \citealt{Strader} show that either the system must be close to edge-on ($i\geq 83^{+7}_{-10}$\,degrees), or the pulsar mass must be higher than 2\,M$_{\odot}$. The system exhibits radio eclipses for
a large fraction of the orbit ($\sim$50\,\%, \citealt{Deneva, Deneva2020}). The radio pulsar has a $\gamma$-ray counterpart, 4FGL J1048.6+2340, which shows $\gamma$-ray pulsations at a statistical significance of $\sim$8$\sigma$ \citep{Deneva2020}. Optical and X-ray studies \citep{Cho, Yap} reveal a strong orbital modulation with rapid variations: the optical light curve changes from an ellipsoidal to a sinusoidal-like profile in less than 14 days. The average optical magnitudes when the pulsar wind heating dominates are $r' \sim 19.4$\,mag and $g' \sim 20.7$\,mag \citep{Yap}. Instead, when the orbital modulation is dominated by ellipsoidal modulation, the system is fainter by 0.3-0.4\,mag. The X-ray light curve is also strongly modulated with a minimum centered at $\phi$ = 0.0, which corresponds to the inferior conjunction of the companion star in our convention. Optical spectra acquired with the 4-m SOAR telescope show the most intense variability yet observed for any RBs \citep{Strader}. The strong emission lines in some instances appear double-peaked, sometimes single-peaked and are absent at other orbital phases. The multiepoch poor orbital phase sampling precludes a secure origin of the emission lines region. 

In this paper we present a multiwavelength campaign of J1048. Using Very Large Telescope (VLT), Sardinia Radio Telescope (SRT), Low-Frequency Array (LOFAR) telescope, Galileo telescope, and {\it Swift} simultaneous observations carried
out during the radio pulsar state, we study the kinematics and geometry of matter in this interacting binary system. In Sect.\,\ref{Obeservations} we describe our data set and data reduction steps. In Sect.\,\ref{Spectrum} we present the average spectrum, the variability of H$\alpha$ emission line and the optical light curve, while the radial velocity curves are presented in Sect.\,\ref{Radial velocity}. In Sect.\,\ref{tomography} we show the H$\alpha$ Doppler tomography. Sects.\,\ref{radio timing}-\ref{optical timing} are dedicated to radio and optical timing analysis. In Sect.\,\ref{discussion}, we discuss the results and we present
future perspectives.
   
\begin{table*}
\renewcommand{\arraystretch}{1.2}
\centering
\caption{Summary of the observations of PSR J1048+2339.}              
\label{table:1}      
\begin{tabular}{l c c c c}          
\hline\hline                        

Telescope/Instrument & Mode & Band & Start time (UTC) &Net exposure (s) \\
\hline
VLT-FORS2 & Imaging &$R$-band & 2020 Mar 19 00:49:07 & 200 \\
VLT-FORS2 & Grism 1200R+93  & 575 - 731 nm &2020 Mar 19 00:59:40 & 19200 \\
SRT-1 & Baseband-Search & P (0.3 - 0.4 GHz) / L (1.3 - 1.8 GHz) & 2019 Aug 26 14:20:28 & 5400 \\
SRT-2 & Baseband-Search & P (0.3 - 0.4 GHz) / L (1.3 - 1.8 GHz) & 2020 Mar 18 20:10:50 & 29960 \\
LOFAR & Raw voltages &110 - 188 MHz & 2020 Mar 18 21:01:00 & 25200\\
\textit{Swift}-XRT &Photon Counting&0.3 - 10 keV & 2020 Mar 18 21:43:14  & 3611 \\
\textit{Swift}-UVOT & Imaging &$U, B, V, UVW1, UVM2, UVW2$ & 	2020 Mar 18 21:43:19  & 3419\\
Galileo/IFI+Iqueye & Fast timing & 400 - 700\,nm  & 2020 Mar 18 23:26:33  & 9000\\

\hline

\end{tabular}
\end{table*}
\section{Observations}
\label{Obeservations}
In March 2020, we performed a multiwavelength campaign focused on J1048 with simultaneous observations in the X-ray, UV, optical and radio bands. Table\,\ref{table:1} lists the observations analyzed and discussed in this
paper. In the following, we describe the data analysis for the different data sets acquired.
\subsection{Optical and ultraviolet observations}
\subsubsection{Very Large Telescope}
Optical spectroscopic observations of J1048 were
carried out on March 19, 2020 at the Observatorio Monte Paranal, using the ESO 8.2\,m Antu Telescope
(UT1) equipped with the FORS2 Spectrograph \citep{Appenzeller1998}. During the night, the seeing ranged between $0.6-0.9''$, with most of the time stable between $0.7-0.8''$. A total of sixteen spectra were obtained, covering a complete orbital cycle, using 1200\,s long exposures. We used a $1''$ slit, always oriented along the parallactic angle, and the grism 1200R+93, centered at 6500~\AA~and with a dispersion of 0.76~\AA/pixel. In order to extract the optical light curve, we acquired one or two $R$-band images with 20~s exposure time, once every two spectroscopic acquisitions, for a total of 10 images.

Image reduction was carried out following standard procedures: subtraction of an averaged bias frame, division by a normalized flat frame. Astrometry was performed using the USNOB1.0 catalog\footnote{\url{http://tdc-www.harvard.edu/catalogs/ub1.html}}. Aperture photometry was made with the PHOTOM software part of the STARLINK\footnote{\url{http://starlink.eao.hawaii.edu/starlink}} package. The photometric calibration was done against Stetson standard stars \citep{Stetson}. In order to minimize any systematic effect, we performed differential photometry with respect to a selection of local isolated and nonsaturated reference stars. The reduction and extraction of the spectra was performed with the ESO-FORS pipeline\footnote{\url{https://www.eso.org/sci/software/pipelines/fors/fors-pipe-recipes.html}} and the ESO-MIDAS\footnote{\url{https://www.eso.org/sci/software/esomidas/}} software package. Wavelength and flux calibration of the spectra were achieved using helium-argon lamp and observing spectrophotometric standard stars, respectively, at the beginning and end of the observing run. We further checked stability of the wavelength calibration using night-sky emission lines.
For the cross-correlation of the spectra with late-type templates and the Doppler tomography analysis we employed the MOLLY package developed by Tom Marsh\footnote{\url{http://deneb.astro.warwick.ac.uk/phsaap/software/}} and the \texttt{pydoppler}\footnote{\url{https://github.com/Alymantara/pydoppler}} package \citep{Spruit1998}, respectively.

\subsubsection{Galileo/IFI+Iqueye telescope}

We observed J1048 with IFI+Iqueye mounted at the 1.2 m Galileo telescope in Asiago, Italy. Iqueye+\footnote{\url{https://web.oapd.inaf.it/zampieri/aqueye-iqueye/index.html}} is a fast photon counter with a field of view of 12.5 arcsec and the capability of time tagging the detected photons with sub-ns time accuracy \citep{Naletto2009}. The instrument is fiber-coupled with the Galileo telescope by means of a dedicated instrument, the Iqueye Fiber Interface (IFI; \citealt{Zampieri2019}). A total of 3 acquisitions (obs. ID 20200319-002631, 20200319-020229, 20200319-032617) were performed on March 18--19, 2020, between 23:26:33.4 and 02:56:18.0 UTC. The first two acquisitions lasted 3600\,s and the third 1800\,s. The sky background was regularly monitored between on-target observations. The average rate at the position of the pulsar varied between $\sim 1300$ and $\sim 1600$\,counts\,s$^{-1}$ because of variations in the sky background. The companion star is not significantly detected.

The data reduction is performed with a dedicated software\footnote{QUEST v. 1.1.5, see \cite{Zampieri2015}.}. The whole acquisition and reduction chain ensures an absolute accuracy of $\sim 0.5$\,ns relative to UTC \citep{Naletto2009}. The photon arrival times are barycentered using TEMPO2 in TDB time units \citep{Hobbs2006} and the JPL ephemerides DE421. The position of the pulsar is RA=10:48:43.4183, DEC=+23:39:53.411 at MJD 56897.0 \citep{Deneva}. We corrected for the motion of the pulsar along the orbit using the orbital period $P_{b} = 0.25051904499812022$\,days and projected semi-major axis $x_p = 0.836122$\,light-seconds of \cite{Deneva}, while the time of passage at the ascending node $\Tasc = 58721.666409$\,MJD (Barycentric Dynamical Time) was derived from a SRT observation on August 19, 2019 (see Sect.\,\ref{radio timing}).

\subsubsection{{\it Swift}-UVOT}
The Ultraviolet Optical Telescope (UVOT) \citep{Roming} on-board the Neil Gehrels \textit{Swift} Observatory \citep{Gehrels} observed J1048 with the $U$ ($\lambda_{c}$= 346.5\,nm), $B$ ($\lambda_{c}$= 439.2\,nm), $V$ ($\lambda_{c}$= 546.8\,nm), $UVW1$ ($\lambda_{c}$= 260.0\,nm), $UVM2$ ($\lambda_{c}$= 224.6\,nm), and $UVW2$ ($\lambda_{c}$=\,192.8\,nm) filters for $\sim$3.4\,ks on March 18--19, 2020 (see Table\,\ref{table:swift}).
We performed aperture photometry by using a circular region centered on the source with a radius of $5''$ for the UV filters and $3''$ for the optical filters. The background emission was extracted from a nearby source-free region. The target was not detected in any of the images. Using the tool \texttt{uvotsource} of the \textit{Swift} analysis software in the HEASoft package version 6.28\footnote{\url{https://heasarc.gsfc.nasa.gov/docs/software/heasoft/}}, we estimated the following 3$\sigma$ upper limits on the optical/UV magnitude: $U>19.9$\,mag, $B>20.3$\,mag, $V>19.3$\,mag, $UVW1>$19.5\,mag, $UVM2>$19.6\,mag, and $UVW2>$20.1\,mag (Vega system).

\begin{table}
\renewcommand{\arraystretch}{1.2}
\centering
\caption{Summary of the {\it Swift}-UVOT observations of PSR\,J1048+2339.}              
\label{table:swift}      
\begin{tabular}{l c c}          
\hline\hline                        

Filter & Start time (UTC) &Net exposure (s) \\
\hline
$UVW1$ & 2020 Mar 18 21:43:19 & 288 \\
 & 2020 Mar 19 03:51:32 & 299 \\
$U$ & 2020 Mar 18 21:46:45& 143 \\
 & 2020 Mar 19 03:53:48 & 149\\
 $B$ &2020 Mar 18 21:48:30& 143 \\
  & 2020 Mar 19 03:54:58 & 149 \\
$UVW2$ &2020 Mar 18 21:50:17& 576 \\
 & 2020 Mar 19 03:56:09 & 600 \\
 $V$ &2020 Mar 18 21:57:07& 143 \\
  & 2020 Mar 19 04:00:39 & 149 \\
  $UVM2$ &2020 Mar 18 21:58:52& 383\\
  & 2020 Mar 19 04:01:49 & 397 \\
\hline

\end{tabular}
\end{table}

\subsection{Radio observations}
\label{sec:radio_observations}

We observed J1048 on two different occasions in the radio band.

On August 19, 2019 we used the 64-m SRT to carry out a 1.5-hour-long observation with the dual-band coaxial L-P receiver \citep{Valente}. In the L band ($\sim1.3-1.8$~GHz), the total-intensity signal was recorded every 125 $\mu$s with the Australia Telescope National Facility (ATNF) Digital Filterbank Mark III backend (DFB\footnote{\url{http://www.jb.man.ac.uk/pulsar/observing/DFB.pdf}}) in search mode. The nominal bandwidth was chosen to be of 1024 MHz, centered at 1548 MHz and divided into 1024 frequency channels. This setup was chosen to prevent aliasing effects within the actual receiver observing band, which is $\sim 500$~MHz wide. The P band was instead acquired in baseband mode with the Reconfigurable Open Architecture Computing Hardware\footnote{\url{http://casper.ssl.berkeley.edu/wiki/ROACH}} (ROACH-1) backend, recording 80 MHz of bandwidth centered at a frequency of 336 MHz. Thanks to the presence of a Gregorian cover, which shields the receiver from radio signals generated by the telescope instrumentation, this observation was mostly free of Radio Frequency Interference (RFI) in both bands.

The second radio observation was carried out in the night between March 18--19, 2020. This time, J1048 was observed simultaneously with the SRT and the Low-Frequency Array telescope (LOFAR, \citealt{Stappers11,vanHaarlem13}), for more than a full orbit.
The SRT observation lasted $\sim8.3$\,hr, and was carried out using exactly the same setup as in the previous observation, with the only difference being that the Gregorian cover was not mounted. This resulted in a higher presence of RFI in these data. LOFAR started the simultaneous observation of J1048 in the VHF band ($110-188$\,MHz) with the high-band antennae of the LOFAR core about 50 minutes later than SRT, and recorded data for 7\,hr. The data were collected as beamformed, 8-bit complex voltages from the two linear polarizations, with a sampling time of 5.12 $\mu$s and divided into 400 frequency sub-bands, with a bandwidth of 0.195 MHz each.

\subsection{X-ray observations}   
We analyzed data sets acquired on March 18--19, 2020 using the \textit{Swift} X-ray Telescope (XRT; \citealt{Burrows05}) configured in the photon counting mode (timing resolution of 2.5\,s; obs. IDs 00034285004 and 00034285005). The source was observed for 3.6\,ks, summing up the two pointings. The source is barely detected, with a signal-to-noise ratio of $S/N\sim2.2$ as evaluated by the \texttt{sosta} tool in the HEASoft package. Its net count rate is (2.0$\pm$0.9)$\times 10^{-3}$\,counts\,s$^{-1}$ over the 0.3--10\,keV energy band (uncertainty at a confidence level of 1$\sigma$). Assuming an absorbed power law model with absorption column density of $N_{\rm H}=2.5\times10^{20}$\,cm$^{-2}$ and photon index of $\Gamma=1.5$ \citep{Cho}, the above count rate translates into an observed flux of $(9\pm4)\times10^{-14}$\,erg\,cm$^{-2}$\,s$^{-1}$ and an intrinsic unabsorbed flux of $(9.5\pm4.0)\times10^{-14}$\,erg\,cm$^{-2}$\,s$^{-1}$, or a luminosity of $(3.3\pm1.4)\times10^{31}$\,erg\,s$^{-1}$
(0.3--10\,keV) for a distance of 1.7\,kpc \citep{Deneva2020}. These values are compatible within the uncertainties with those derived by \cite{Cho} and \cite{Yap} using {\em Chandra} deeper observations, and indicate 
that, at the epoch of our multiband campaign, the system remained in a rotation-powered radio pulsar state.

\section{Average spectrum and orbital variability}
\label{Spectrum}
\begin{figure*}
   \centering
   \resizebox{\hsize}{!}{\includegraphics{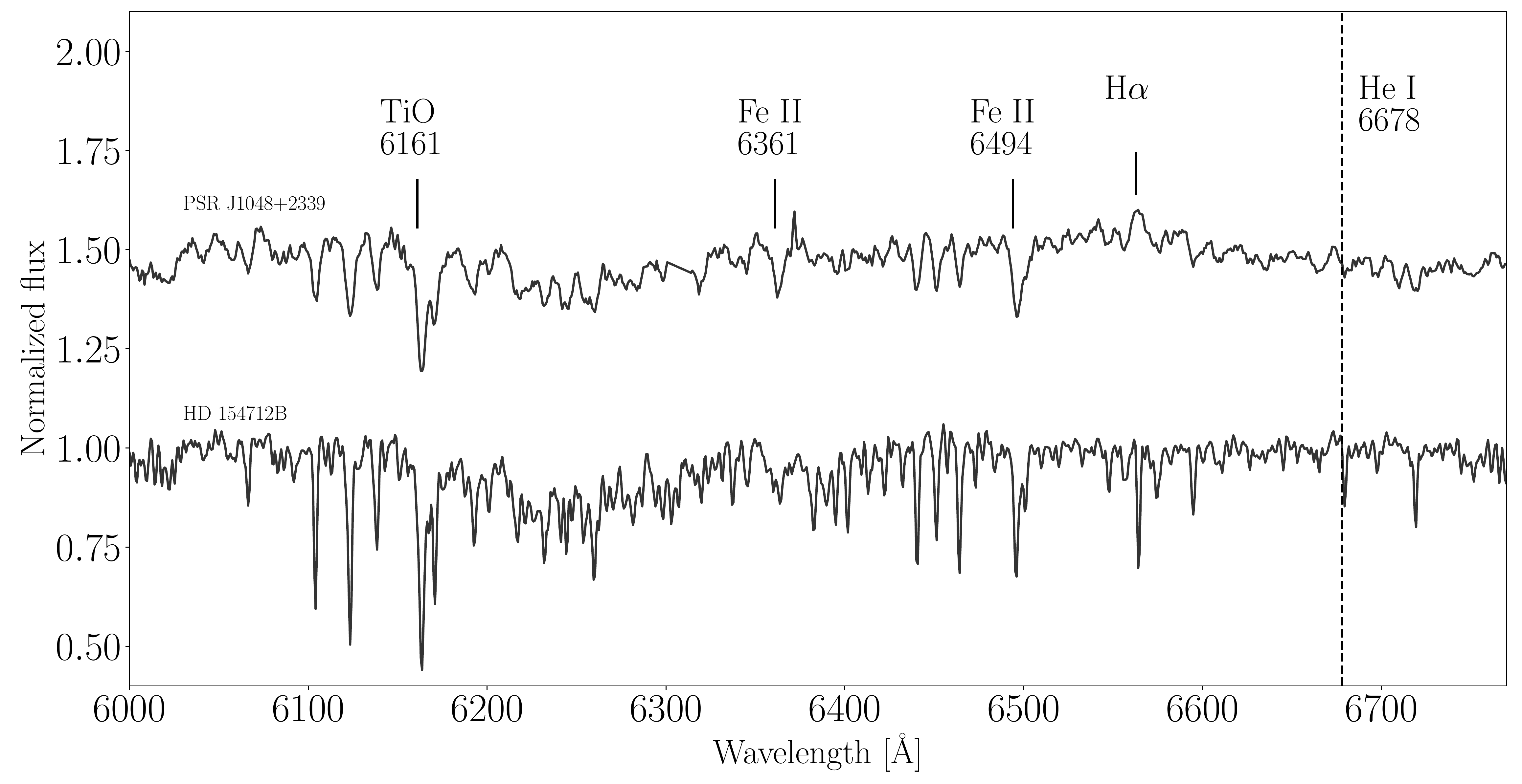}}
   \centering
      \caption{Average optical spectrum of PSR J1048+2339 (top spectrum) compared with a K8 V template (HD 154712B, bottom spectrum). Arbitrary vertical offsets have been applied for clarity. Main emission and absorption lines are indicated. The vertical black dashed line represents the rest wavelength of He I $\lambda$6678, which is absent in our spectra.}
         \label{Fig_av_spec}
   \end{figure*}

The average optical spectrum of J1048 is dominated by absorption lines typical of late-type stars (for instance TiO\,$\lambda$6161, FeII\,$\lambda$6361 and FeII\,$\lambda$6494) and by the H$\alpha$ emission line. It is displayed in Fig.\,\ref{Fig_av_spec} together with a K8\,V template, HD\,154712B, that well matches the absorption features (see also Sect.\,\ref{Radial velocity}). One of the typical features of the accretion disk, the HeI $\lambda6678$ emission line, is absent. The H$\alpha$ profile in the average spectrum is complex and appears single-peaked, casting doubts on the presence of an accretion disk.

\begin{figure*}
   \centering
   \resizebox{\hsize}{!}{\includegraphics{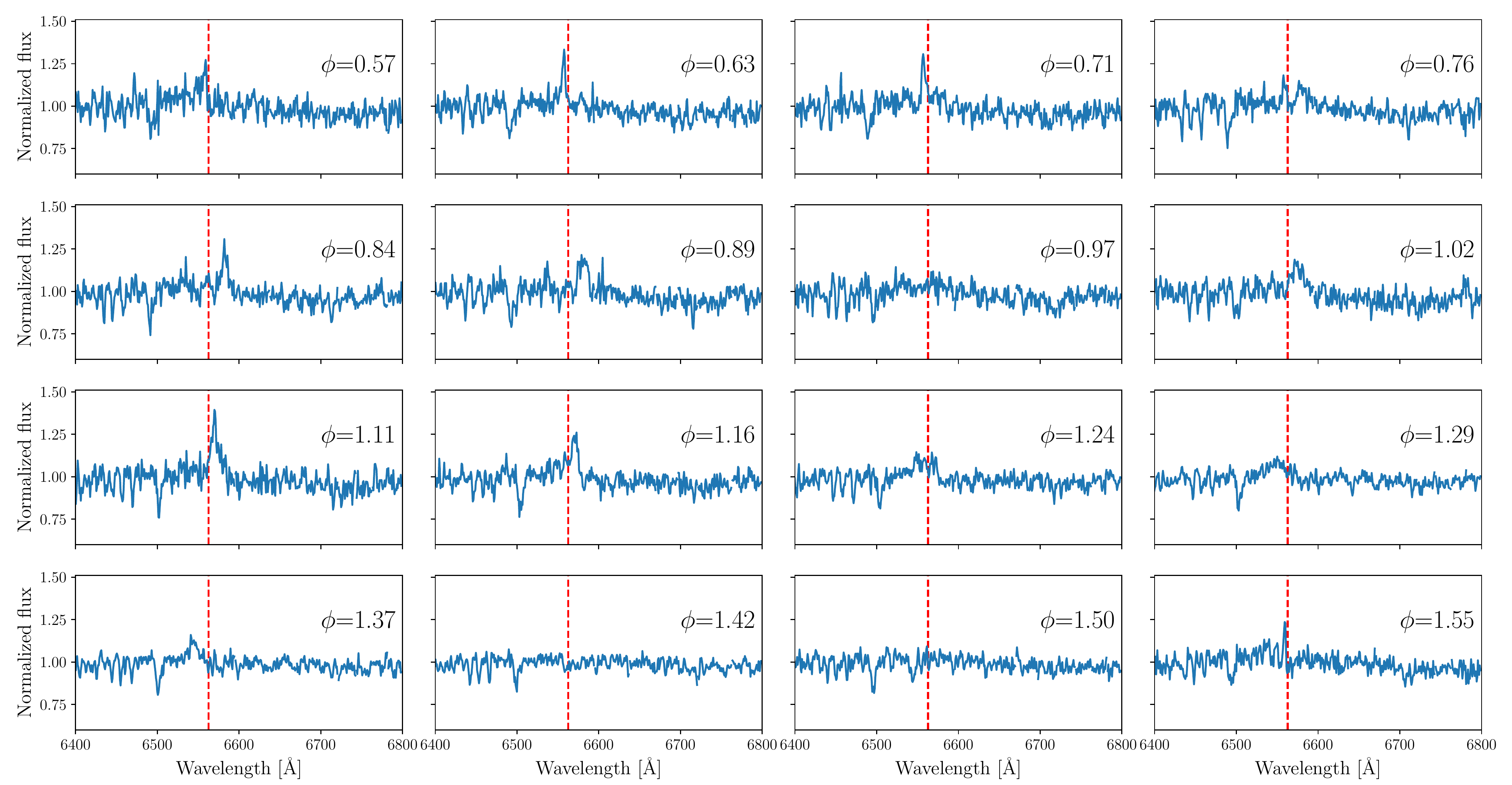}}
   \centering
      \caption{H$\alpha$ emission variability along the whole orbit. The orbit phase increases from left to right, from top to bottom. The vertical red dashed line represents the H$\alpha\,\lambda$6561 rest wavelength. We use the convention where $\phi$=0 is the point where the companion star is closer to the observer. 
      }
         \label{Fig_Ha_variability}
   \end{figure*}

The H$\alpha$ emission shows strong variations along the orbit (Fig.\,\ref{Fig_Ha_variability}): at some orbital phases it disappears (for example at $\phi$= 0.42), while at other epochs it appears with a single component ($\phi$=0.63) and sometimes it displays a weak secondary peak ($\phi$=0.84). 
Binary phases are computed using the extremely accurate radio pulsar ephemeris (see Sect.\,\ref{radio timing}). Given the nearly circular orbit, the choice of the time of the ascending node $\Tasc$ is arbitrary and in the paper we opted to decrement that by $P_b/4$, where $P_b$ is the orbital period, with respect to the one computed from radio ephemeris, so that phase 0 corresponds to the inferior conjunction of the companion star.

\begin{figure*}
   \centering
   \includegraphics[width=11cm]{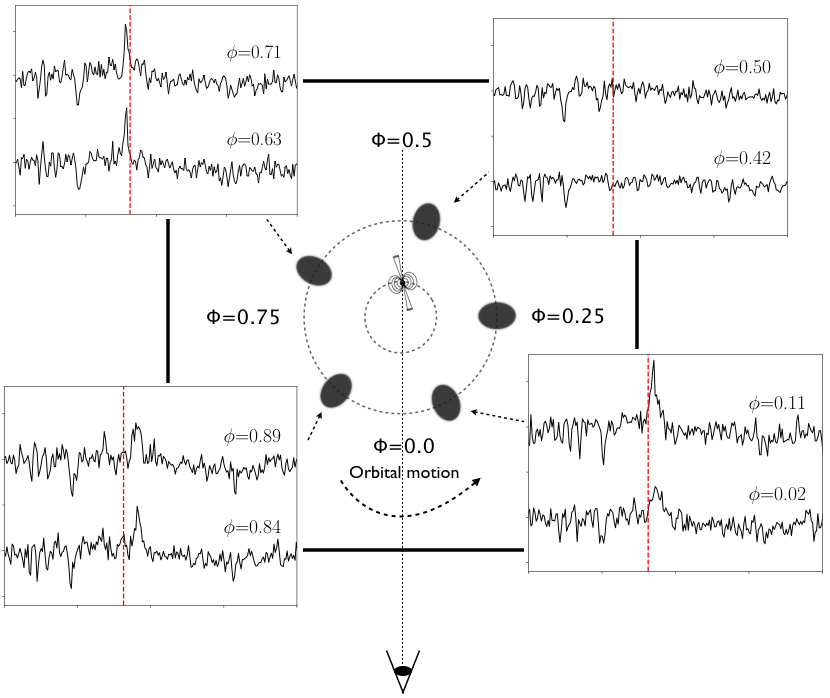}
   \includegraphics[width=7cm]{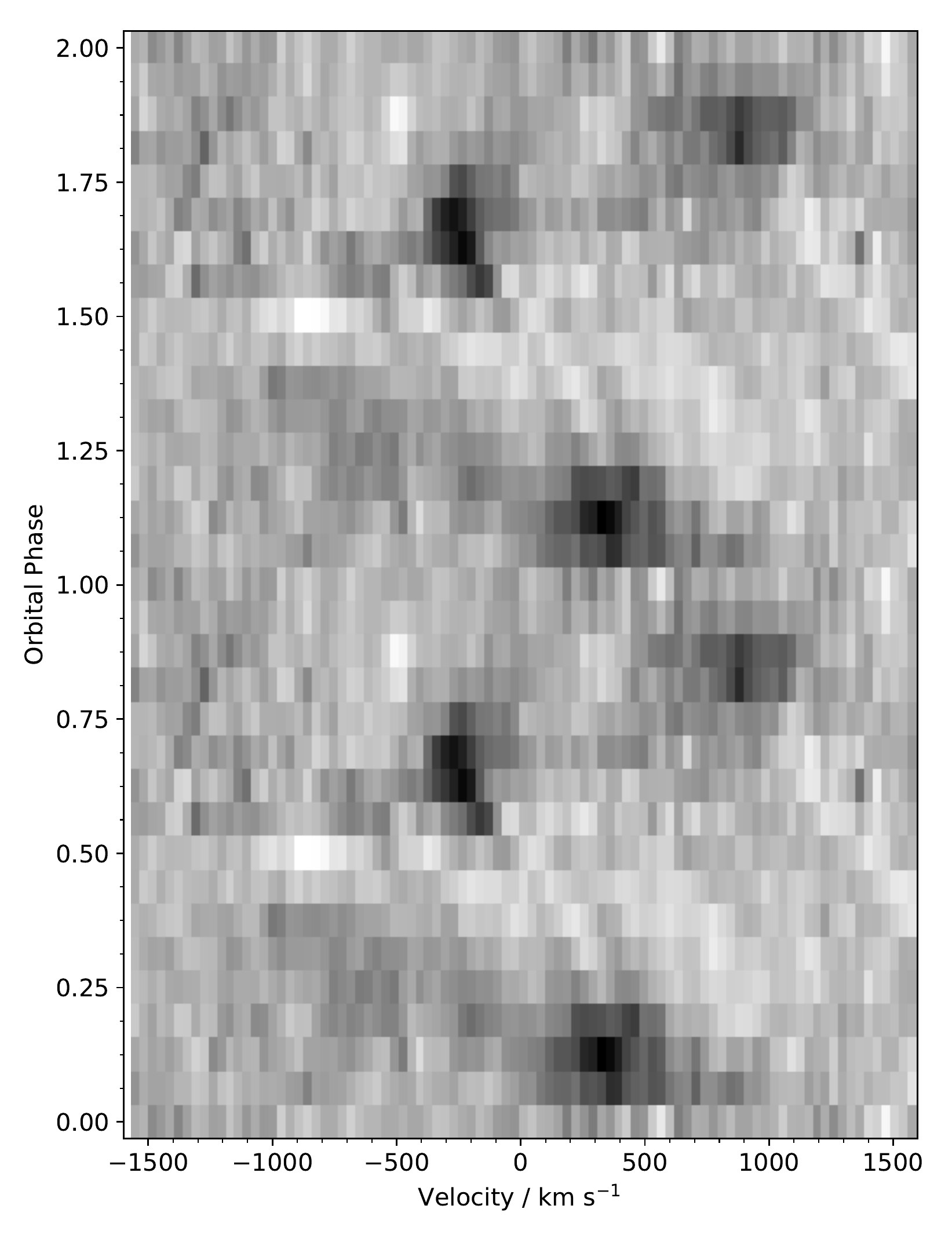}
   
      \caption{Left: A schematic panel of PSR J1048+2339 showing the orbital motion of the companion with H$\alpha$ variability. The vertical red dashed line represents the H$\alpha\,\lambda$6561 rest wavelength. Right: trailed spectrum showing the orbital evolution of the H$\alpha$ emission line. The spectra have been normalized to the continuum. Two cycles are shown for clarity. 
              }
         \label{Fig_Ha}
   \end{figure*}

Differently from what was found by \cite{Strader}, in our case the H$\alpha$ emission is weak and asymmetric, with stronger blue or red components depending on the orbital phase. Near the superior conjunction ($\phi$=0.4--0.5), the H$\alpha$ emission has almost completely vanished or is weakly detected as an absorption feature, while near the inferior conjunction ($\phi$=0) two peaks emerge (see Fig.\,\ref{Fig_Ha}, left plot). The blue component is weaker and associated with the companion star that is approaching the Earth. The red component vice versa is emitted by material receding from the Earth along our line-of-sight. This could be ascribed to the material flowing from the donor star toward the pulsar. We constructed the trailed spectrogram, shown in the right panel of Fig.\,\ref{Fig_Ha}, to investigate the orbital line variability in more detail. It displays the orbital evolution of the H$\alpha$ emission line in 16 phase bins. Between phases 0.75 and 0.90, an emission spot with positive velocities that does not trace the orbital motion of the companion star is clearly visible. This is a first indication for the presence of material overflowing from the Roche lobe, confirmed by the Doppler tomography study (see Sect.\,\ref{tomography}). The gas from which the H$\alpha$ emission originates together with the irradiated companion star are probably responsible for the radio eclipses of J1048 that cover about half of the orbit \citep{Deneva, Deneva2020}. 

\begin{figure}
   \centering
   \resizebox{\hsize}{!}{\includegraphics{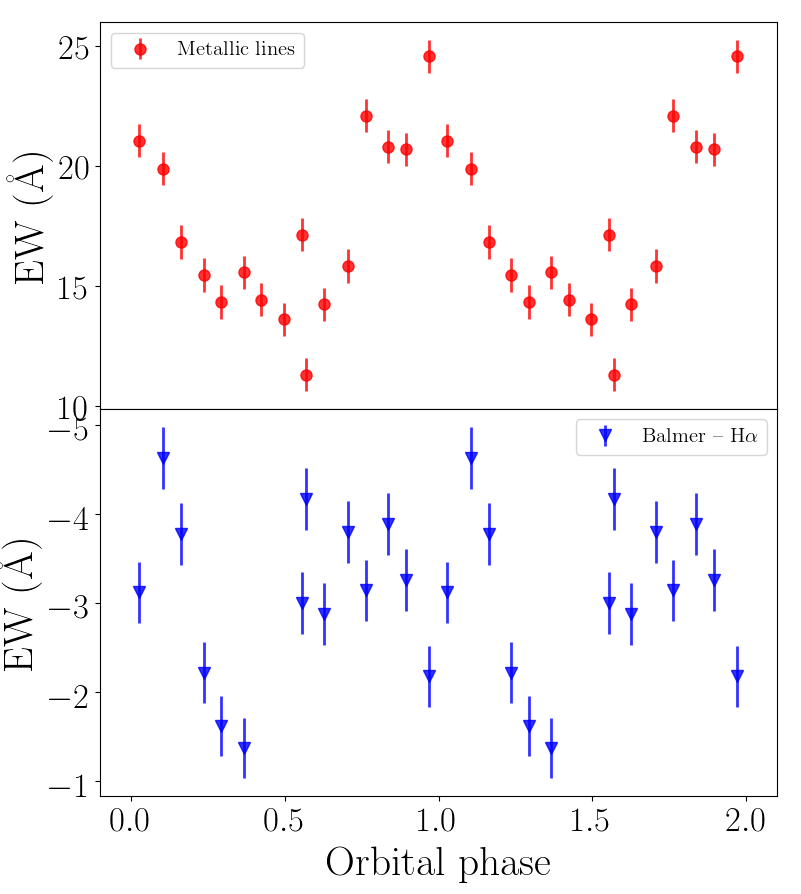}}
   \centering
      \caption{Equivalent width of metallic (red dots) and H$\alpha$ (blue triangles) absorption and emission lines, respectively, as a function of the orbital phase. Two cycles are shown for clarity.}
         \label{Fig_Ha_ew}
   \end{figure}

Measurements of the equivalent widths (EWs) of the H$\alpha$ emission line show a hint of modulation at the orbital period although at low statistical significance ($\sim 1.9\sigma$, see Fig.\,\ref{Fig_Ha_ew}, bottom panel). The mean EW of H$\alpha$ is -3.1$\pm$1.3\,\AA. In the trailed spectra of H$\alpha$ in Fig.\,\ref{Fig_Ha} the emission is at maximum at orbital phases 0.1--0.2 and 0.6--0.7. For comparison, we measured the EWs of metallic lines (in the spectral ranges 5970--6291\,\AA\,and 6421--6522\,\AA). The EW curve of metallic lines shows almost a sinusoidal profile with a single peak significance at a 3.9$\sigma$, while the EW curve of H$\alpha$ has a noisy profile. This might be due to the fact that the H$\alpha$ line is contaminated by extended emission as hinted at by the trailed spectra (Fig.\,\ref{Fig_Ha}), while the metallic lines originate only from the companion star. 

\begin{figure}
   \centering
   \resizebox{\hsize}{!}{\includegraphics{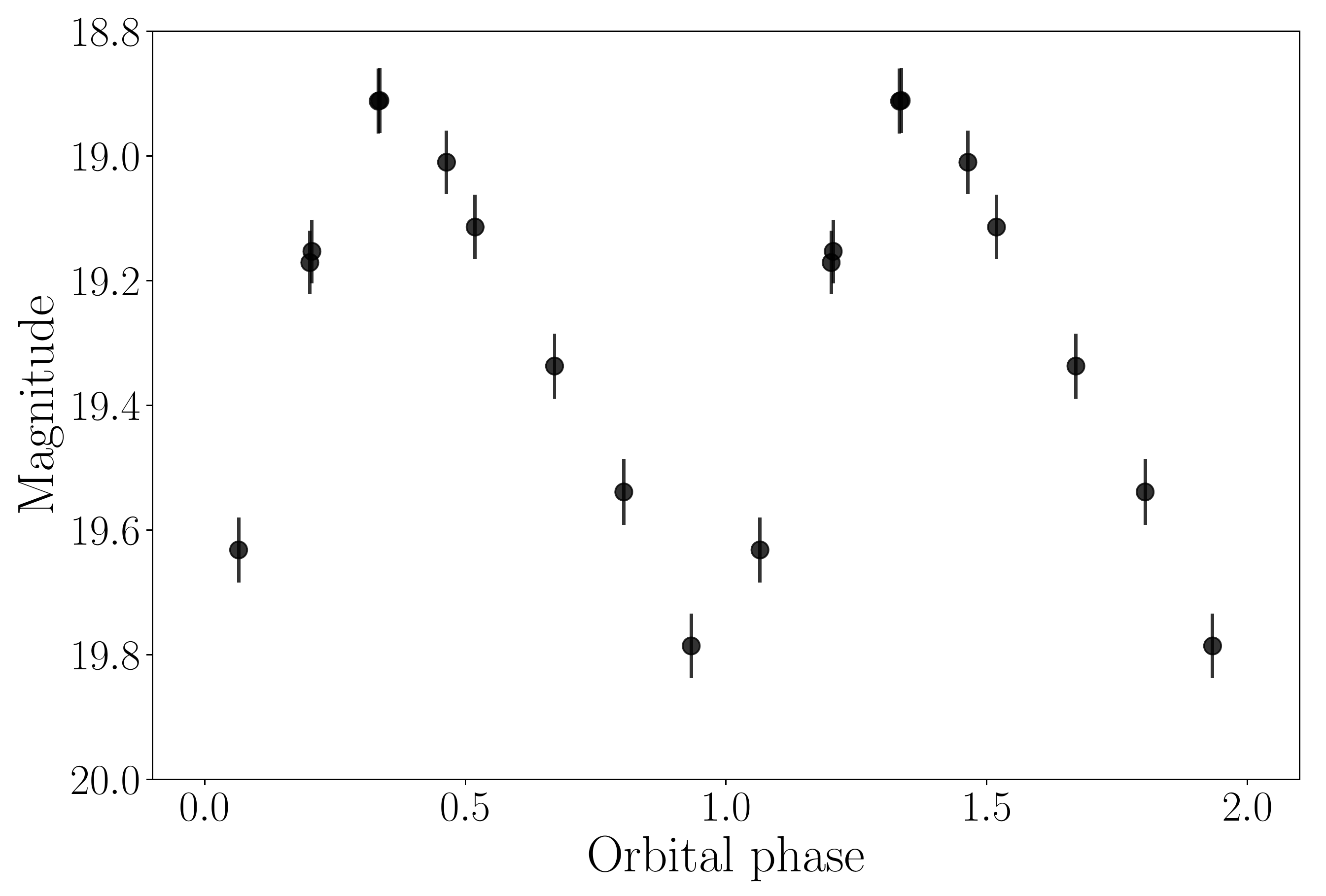}}
   \centering
      \caption{Phase-folded $R$-band light curve of PSR J1048+2339 companion star. The light curve shows a sinusoidal-like modulation with a peak-to-peak amplitude of $\sim$0.9\,mag. The minimum is around phase zero as expected, while the maximum slightly precedes phase 0.5. Two cycles are shown for clarity. }
         \label{Fig_LC}
   \end{figure}

The simultaneous $R$-band light curve (Fig.\,\ref{Fig_LC}) shows an asymmetric sinusoidal-like profile with a minimum around $\phi=~0$ and an average magnitude $R\sim19.3$\,mag (Vega). The peak-to-peak amplitude of the light curve is of about 0.9 magnitude. The light curve shape and the magnitude variations along the orbit suggest that the optical emission of the system is dominated by the emission from the hot face of the companion, heated by the pulsar wind.
The curve does not have a pure sinusoidal profile, the minimum is around phase zero as expected, while the maximum precedes phase 0.5. 

\begin{figure}
   \centering
   \resizebox{\hsize}{!}{\includegraphics{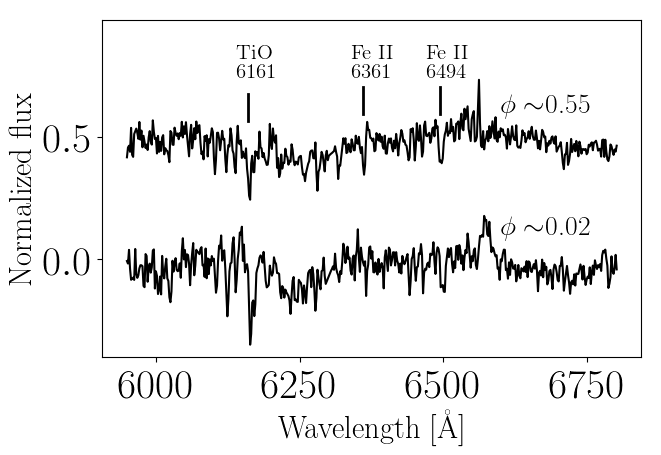}}
   \centering
      \caption{Main metallic absorption lines as observed at the inferior conjunction ($\phi \sim 0.02$) and at the superior conjunction ($\phi \sim0.55$).}
         \label{Fig_righe_metalliche}
   \end{figure}

Differently from  what \cite{Linares2018} observed in the RB PSR\,J2215+5135, we do not find drastic changes in the spectral absorption lines of the irradiated and the cold face of the companion star (Fig.\,\ref{Fig_righe_metalliche}). In PSR\,J2215+5135, Balmer and Mg-I triplet absorption lines show strong EW variation as a function of the orbital phase due to the irradiation by the pulsar. The EWs of Balmer and  Mg-I triplet absorption lines change by a factor $\simeq 7$ and $\simeq 3$, respectively \citep{Linares2018}. Indeed, optical photometry of PSR\,J2215+5135 revealed single humped modulation with magnitude variations of $\sim1.5$\,mag along the orbit, typical of strongly irradiated system \citep{Breton2013}. In J1048, the EWs of H$\alpha$ emission line and of metallic absorption lines vary by a small factor ($\simeq3$ and  $\simeq2$ respectively) along the orbit and we do not observe large temperature changes between the dark and bright sides of the companion star. The $\sim0.9$\,mag variation in amplitude observed in the $R$-band light curve could therefore be attributed to a moderate pulsar irradiation.

\section{Radial velocity}
\label{Radial velocity}
Following \cite{Yap} we cross-correlated the 16 VLT-FORS2 spectra with six K\,V template stars in the spectral ranges 5970--6291\,\AA\,and 6421--6522\,\AA\ (taken from \citealt{Casares1996}), masking the main telluric and interstellar features and the H$\alpha$ emission line, obtaining six radial velocity curves for each template star. We then performed a (least-squares) sine fit to our radial velocity data, allowing the projected radial velocity semi-amplitude ($K_2$) and the systemic velocity of the companion star ($\gamma$) to vary. We rescaled the error bars of the FORS2 measurements by a factor of 4.3, in order to obtain a reduced $\chi^2=1$. The best match was obtained for the K2\,V and K8\,V templates, with equal significance (see Table\,\ref{tab:radial velocity}).

\begin{table*}
\centering
\renewcommand{\arraystretch}{1.2}                                      
\caption{Radial velocity parameters of PSR J1048+2339 with estimates of 1$\sigma$/68\% uncertainties. We rescaled the error bars of the VLT measurements by a factor of 4.3 to obtain a reduced $\chi^2$=1.}
\label{table:2}      
\begin{tabular}{c c c c c c}          
\hline
\hline                        
Template star& Spectral Type & $\gamma $ & K$_2$ & $\chi^2$/d.o.f & d.o.f\\
 & & (km s$^{-1}$) & (km s$^{-1}$) & \\  
\hline 
HD~184467 &K2 V &-14.4$\pm$3.7 &339.9$\pm$5.4 &1.0 &13 \\
HD~29697 &K3 V &-17.8$\pm$3.2 &341.0$\pm$4.5 &1.2&13 \\
HD~154712A &K4 V&-10.2$\pm$3.3 &341.3$\pm$4.7   &1.4&13 \\
61 Cyg A & K5 V &-17.9$\pm$3.1 & 341.2$\pm$4.4&1.4&13 \\
61 Cyg B & K7 V &-17.2$\pm$3.1 & 342.3$\pm$4.4&1.3&13 \\
HD~154712B & K8 V&-10.5$\pm$3.1 &343.3$\pm$4.4  &1.0&13 \\
\hline

\\
\end{tabular}
\label{tab:radial velocity}
\end{table*}

We employed two independent approaches to classify the spectral type of the companion star. In the first one, we compared the ratio of the EW of the absorption lines in J1048 with different stellar templates. We considered the ratios $\lambda$6361(FeII)/$\lambda$6161 (TiO) and $\lambda$6494(FeII)/$\lambda$6161 (TiO). Using this criterion, the most likely spectral types are K7\,V-K8\,V (see Table\,\ref{tab:lines ratio}).

 \begin{table}[htb]
\renewcommand{\arraystretch}{1.2}
\caption{Line ratios for PSR J1048+2339 and template stars.}
\label{table:ratios}      
\begin{tabular}{c c c c}          
\hline
\hline                        
Template star& Spectral Type & 6361(Fe\,II)/ & 6494(Fe\,II)/\\
 & & 6161(TiO)& 6161(TiO)  \\  
\hline 
PSR J1048 & &0.22$\pm$0.01&0.34$\pm$0.01  \\
\hline
HD~184467 &K2 V &0.36$\pm$0.01 &0.56$\pm$0.01 \\
HD~29697 &K3 V &0.27$\pm$0.01 &0.52$\pm$0.01 \\
HD~154712A &K4 V&0.40$\pm$0.01 &0.57$\pm$0.01  \\
61 Cyg A & K5 V &0.27$\pm$0.01 & 0.45$\pm$0.01 \\
61 Cyg B & K7 V &0.30$\pm$0.01 & 0.33$\pm$0.01 \\
HD~154712B &K8 V &0.26$\pm$0.01 &0.36$\pm$0.01 \\
\hline

\\
\end{tabular}
\label{tab:lines ratio}
\end{table}

In an alternative approach, we subtracted different templates appropriately degraded from the target average spectrum.  We broadened
stellar templates from 10 to 200 km\,s$^{-1}$, in steps of
10 km\,s$^{-1}$ and subtracted the resultant broadened spectra
from the J1048 Doppler-corrected average spectrum in the spectral range 6432-6477\,\AA, using a limb darkening coefficient $\mu$=0.5. This subtraction method allowed us to estimate the companion star rotational broadening $v\,{\rm sin}\,i$. In the optimal subtraction, the stellar template is scaled by a factor $0 \leqslant f_{frac} \leqslant 1$, which represents the fraction of flux from the companion star. We used this quantitative approach to match the observed absorption lines from
J1048 to set of templates used before. The best fit was obtained for the template K8\,V with a star contribution of $f_{frac}=0.56\pm$0.02 and a projected rotational velocity for the companion of $v\,{\rm sin}\,i$= 105$\pm$15\,km\,s$^{-1}$ (see Table\,\ref{tab:spectral type}). We verified that this result is not significantly affected by the variation of the limb darkening $ \mu$ in the range of $\left[  0.0, 0.8 \right]$. 

 \begin{table}[htb]
\renewcommand{\arraystretch}{1.2}
\caption{PSR J1048+2339 spectral classification by direct fitting using the limb darkening coefficient $\mu$=0.5 and the rotational broadening $v\,{\rm sin}\,i$\,=\,105$\pm$15\,km\,s$^{-1}$.}     
\begin{tabular}{c c c c}          
\hline
\hline                        
Template star& Spectral Type & $f_{frac}$ & $\chi^2$/d.o.f \\
 & & & (59 d.o.f.)  \\  
\hline
HD~184467 &K2 V &1.06$\pm$0.06 & 3.2 \\
HD~29697 &K3 V &0.64$\pm$0.02 &2.9  \\
HD~154712A &K4 V&0.75$\pm$0.02 &2.4  \\
61 Cyg A & K5 V &0.67$\pm$0.02 &2.6 \\
61 Cyg B & K7 V &0.60$\pm$0.02 &2.9  \\
HD~154712B &K8 V &0.56$\pm$0.02 &2.3  \\
\hline

\\
\end{tabular}
\label{tab:spectral type}

\end{table}

\begin{figure}
   \centering
   \includegraphics[width=9cm]{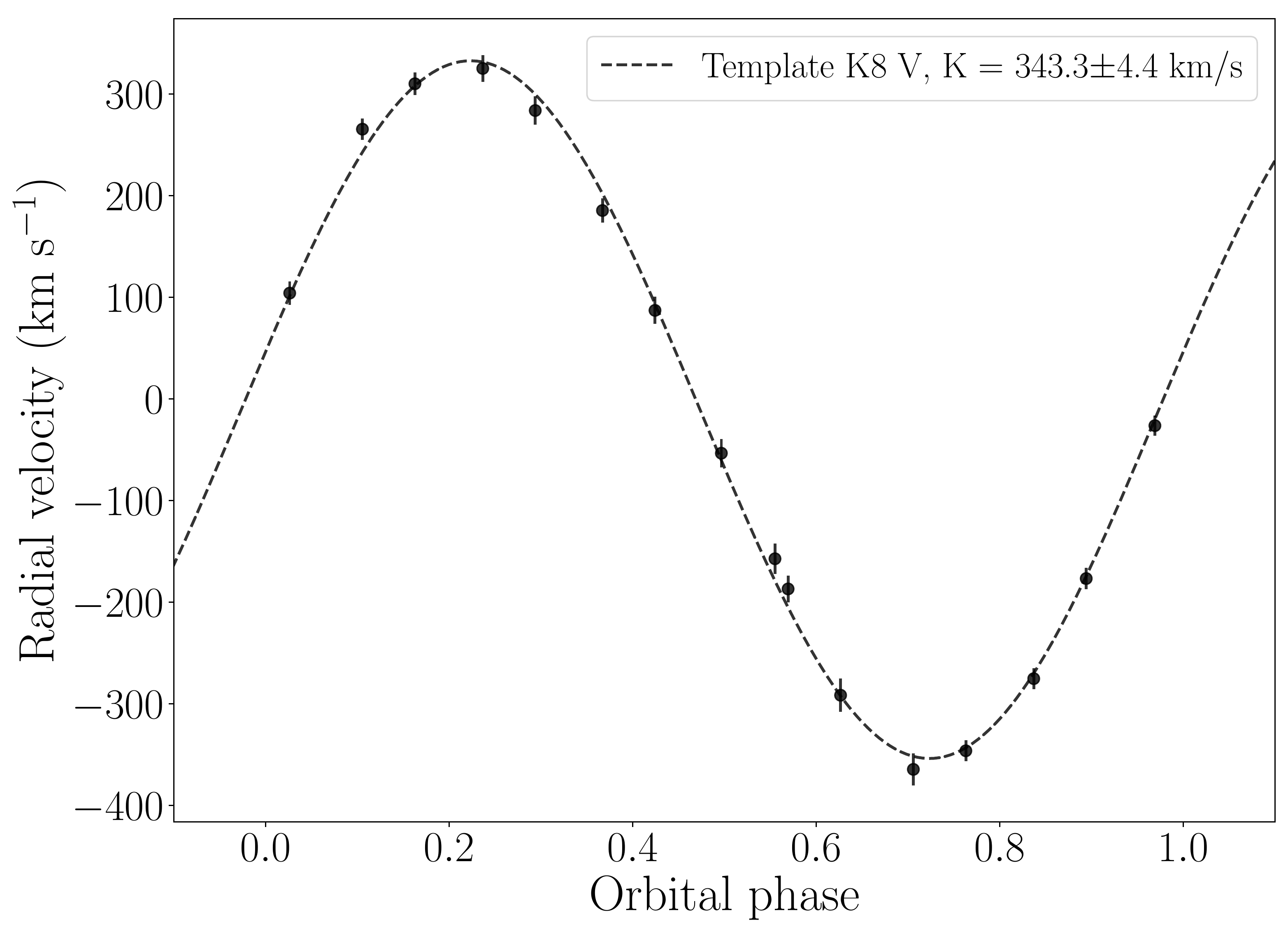}
   \centering
      \caption{Radial velocities of the companion star in PSR J1048+2339 computed using a K8\,V template and folded on the ephemeris derived using SRT. The best-fitting sinusoidal radial velocity curve is superposed. }
         \label{fig_rv}
   \end{figure}

Based on the results of these three analyses, we conclude that the most
likely spectral type for J1048 is a K8\,V star ($T_{eff}\sim4000$\,K). This determination is consistent with what was found by \cite{Yap} and has an uncertain of a couple of subspectral types. We found $K_{\rm 2,obs}$=343.3$\pm$4.4\,km\,s$^{-1}$ and $\gamma$=\,-10.5$\pm$3.1\,km\,s$^{-1}$ (see Table \ref{table:2} and Fig.\,\ref{fig_rv}). However, our constraint on $K_{\rm 2,obs}$ does not seem to depend on the fine determination of the spectral type. 

\subsection{System parameters}

Our determination of the mass ratio $q$ and the NS mass relies on the determination of $K_2$ obtained by the measurement of the Doppler shifts of the absorption lines along the orbital period. This measurement is affected by a systematic error due to the pulsar irradiation, which determines a nonuniform absorption distribution on the stellar surface. The absorption lines formed on the companion photosphere are partially quenched in its irradiated face so that the center of light of these lines is shifted toward the outer face of the companion, causing a systematic shift in the velocity determined at each binary phase. Consequently, the radial velocity curve is not exactly sinusoidal and the $K_2$ obtained by fitting a sine curve is different from that corresponding to the center of mass of the donor. In our case, considering only absorption lines in the cross-correlation with stellar templates, the observed $K_2$ can be overestimated. 

As shown in \cite{Wade1988}, we corrected the velocity using the so-called $K$-correction. The corrected observed radial velocity semi–amplitude is:
\begin{equation}
 K_{\rm 2,corr}(q)=K_{\rm 2,obs}-\Delta K(q)= K_{\rm 2,obs}-f_K(q)\,v\,{\rm sin}\,i.
 \label{eq1}
\end{equation}
Here, $f_K$ is a geometrical correction factor smaller than unity ($0<f_K<1$).
$f_K=0$ indicates uniform emission over the donor star and thus no $K$-correction. 
If the nonirradiated hemisphere has a uniform absorption line strength and the other hemisphere does not contribute to
the observed stellar absorption lines at all, $f_K=4/3\pi \approx 0.42$ \citep{Wade1988}. 
Hence, applying the $K$-correction (Equation\,\ref{eq1}), we get $K_{\rm 2,corr}$=298.7$\pm$7.7\,km\,s$^{-1}$. This is an unrealistic assumption in the case of J1048 that leads to obtain a firm lower limit of $K_2$, because both hemispheres contribute to the observed absorption lines (see Fig.\,\ref{Fig_righe_metalliche}). On the other hand, the measured value of $K_2=343.3 \pm 4.4$\,km\,s$^{-1}$, without $K$-correction, is an upper limit.


Using the NS mass function from \cite{Deneva}, we derived the semi-amplitude of the pulsar projected radial velocity $K_1=(2\pi\,c\,x_p)/P_b$=72.7396$\pm$0.0003\,km\,s$^{-1}$, where $x_p$ is the projected semi-major axis of the pulsar orbit and $P_b$ is the orbital period. By combining $K_1$ and $K_2$ ($291 < K_2 < 348$\,km\,s$^{-1}$, at 1$\sigma$\,c.l.), we found a mass ratio $q=M_2/M_{NS}= K_1/K_{\rm 2,corr}$ of $0.209 < q < 0.250$. This implies that the NS mass $M_{\rm NS}$=$P_b\,K_2^3\,(1+q)^2/(2 \pi\,G\,{\rm sin}^3i)$ is $(1.0 < M_{\rm NS} < 1.6)\,{\rm sin}^{-3}i\,M_{\odot}$. Fixing the inclination to the upper limit of $76^{\circ}$ due to the absence of X-ray eclipses \citep{Yap}, we constrained the NS mass to the range of $1.08< M_{\rm NS} < 1.76\,M_{\odot}$ and the mass of the companion $M_2=(K_1\,M_{\rm NS})/K_2$ to the range of $0.27< M_2 < 0.37\,M_{\odot}$. We favor values of $M_{\rm NS}$ closer to the more massive scenario, because the irradiation seems to weakly affect the $K_2$ measurement. On the other hand, assuming the NS mass upper limit of $\sim 2.3\,M_{\odot}$ \citep{Margalit2017}, we can derive a firm lower limit inclination of $62^{\circ}$ for a nonirradiated system ($f_K=0$) and of $53^{\circ}$ for a strongly irradiated system ($f_K=4/3\,\pi$).

We note that the values of $K_2$, $M_{\rm NS}$ and $M_2$ obtained here are systematically lower than those obtained from \cite{Strader}. However, the data of \cite{Strader} are spread over four nonconsecutive nights (whereas our data were collected over the same night) and they are not adjusted for the $K$-correction. Without the $K$-correction, we get an upper limit of $M_{\rm NS}\leqslant $1.76\,$M_{\odot}$ (at 1$\sigma$\,c.l.), compatible within the errors with the value of \cite{Strader}.

\section{H$\alpha$ Doppler tomography}
\label{tomography}
\begin{figure*}
   \centering
   \resizebox{\hsize}{!}{\includegraphics{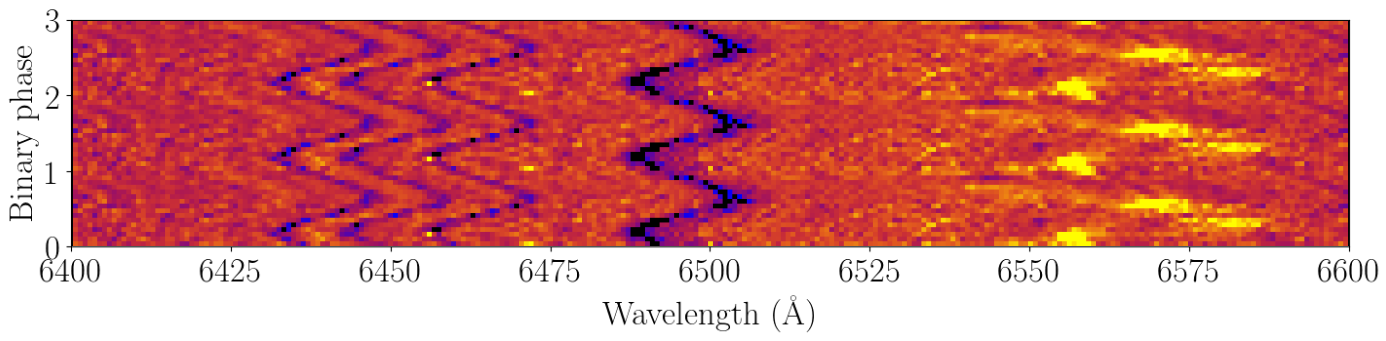}}
      \caption{Trailed spectra averaged in 16 orbital phase bins revealing the orbital evolution of several absorption features and H$\alpha$ emission lines. Three orbital phases are shown for clarify.}
         \label{Fig_trailed_all}
   \end{figure*}

\begin{figure}
   \centering
   \resizebox{\hsize}{!}{\includegraphics{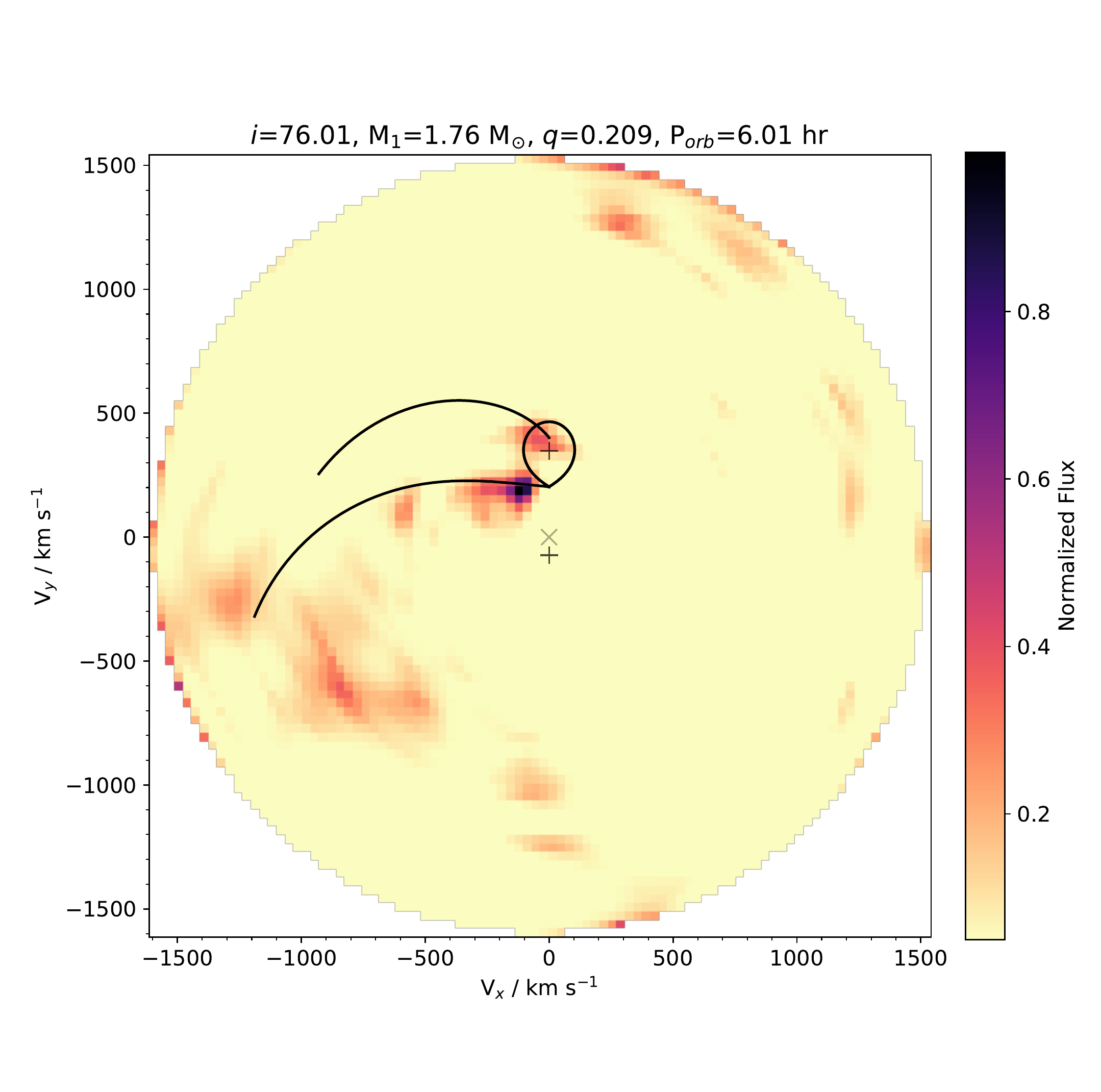}}
   \caption{Doppler map of the H$\alpha$ emission lines. We plot the gravitational free-fall gas stream trajectory (lower curve), the Keplerian disk velocity along the stream (upper curve) and the Roche lobe of the donor star using $K_1$=72.7639\,km\,s$^{-1}$ \citep{Deneva}, $i=76^{\circ}$ \citep{Strader, Yap}, $M_{\rm NS}=1.76$\,$M_{\odot}$ and $q=0.209$. The plus signs represent, from top to bottom, the companion and the NS, while the `$\times$' sign represents the center of mass. The map was computed for a systemic velocity $\gamma=-10.5$\,km\,s$^{-1}$.}
         \label{Fig_dopplermap}
   \end{figure}

We constructed a trailed spectrogram covering the 6250--6700\,\AA~interval to compare the absorption and emission lines structures (see Fig.\,\ref{Fig_trailed_all}). The absorption features are visible throughout the orbital cycle and trace the companion star motion, while the H$\alpha$ emission lines fade at some orbital phases (see Sect.\,\ref{Spectrum}). 
The derived Doppler image of the H$\alpha$ emission line computed with the maximum entropy method \citep{Marsh1988} is shown in velocity coordinates in Fig.\,\ref{Fig_dopplermap}. The Roche lobe of the companion and the gravitational free-fall path of the gas stream are superposed over the map. Fig.\,\ref{Fig_dopplermap} shows the map constructed assuming an inclination $i\sim 76^{\circ}$ and using the $K_2$ upper limit with $M_{\rm NS}\leqslant 1.76\,M_{\odot}$ and $q\geqslant0.209$.
 
This image cannot be interpreted as a time-average map because the data were acquired over one night and the system displays strong variability on very short timescale (few days; \citealt{Wang, Yap}). Therefore, this Doppler map represents the 2D velocity distributions of the H$\alpha$ emission on March 19, 2020. 

The map does not show evidence of an accretion disk, but shows a relatively faint spot emission coincident in phase and velocity with the companion star. The S-wave profile component from the companion is detected in the trailed spectra (see Fig.\,\ref{Fig_Ha}), although the emission is not continuous along the orbit. In addition, the H$\alpha$ map reveals an extended bright spot in the upper left velocity quadrant close to the companion star, suggesting an intra-binary shock created from the interaction between the pulsar relativistic wind and matter leaving the companion star. The proximity of the intra-binary shock to the companion star indicates  the presence of material in that region. If the secondary is filling its Roche-lobe the material streams away from L$_1$, otherwise the companion mass loss mechanism is driven by the ablation of the outer layers of the star by the energetic pulsar wind. Given the orbital separation and the mass ratio $q$, we can determine the approximate Roche-lobe radius $R_L$ \citep{Eggleton1983}. Assuming $i\sim76^{\circ}$ we obtain $0.50\,R_{\odot}< R_L< 0.55\,R_{\odot}$, comparable to the radius of a K8 V star \citep{Pasinetti}. \cite{Yap} derived a Roche-lobe filling factor of 0.86$\pm0.02$ during a phase of single-peaked optical light curve, casting doubts on whether the secondary is truly filling its Roche-lobe. Given the uncertainty in $K_2$ and in the dimension of the companion star irradiated by the pulsar wind, our analysis does not favor one mechanism for the mass loss over the other. There is no evidence for emission from an accretion disk, although its formation could take place on a very rapid timescale, as in IGR J18245--2452 \citep{Papitto2013}.

\section{Radio timing analysis}
\label{radio timing}

The raw data taken in baseband mode with the SRT in the P band, and with the LOFAR in the VHF band, were first coherently dedispersed \citep{Bassa2017} at J1048's nominal dispersion measure (DM) of 16.65~\dmunit. The four Stokes parameters were summed to form total-intensity search-mode files.
For the SRT P-band data, these retained a time resolution of 64 $\mu$s and 640, 125-KHz-wide frequency channels. For the LOFAR VHF-band data, the search-mode files kept a time resolution of 81.92 $\mu$s and 1600, 48-kHz-wide channels.
As explained in Section \ref{sec:radio_observations}, the SRT L-band data were already taken in search mode, so no prior manipulation was needed for them.

Once all the observations were available as search-mode files, we conducted the analysis as follows. For each observation, we first created a mask to filter out the strongest RFI present in the data, using the \texttt{rfifind} routine, part of the \texttt{PRESTO}\footnote{\url{https://www.cv.nrao.edu/~sransom/presto}} pulsar searching package \citep{Ransom+2002}. Taking the mask into account, we then used \texttt{PRESTO}'s \texttt{prepdata} to generate an RFI-free, frequency-summed dedispersed time series. These were folded using \texttt{PRESTO}'s \texttt{prepfold} and the ephemeris published by \cite{Deneva}. However, it is well known that many RBs and BWs show a strong orbital variability \citep{Shaifullah+2016,BakNielsen+2020,Hebbar+2020}, often causing an old ephemeris not to be able to predict the actual orbital phase at a future time. This results in a wrong correction for the orbital R{\o}mer delay \citep{Blandford} and, consequently, the pulsed emission can be missed when folding.
Following \cite{Ridolfi}, we overcame this issue by doing a brute-force search in the time of the ascending node, $\Tasc$, using \texttt{SPIDER\_TWISTER}\footnote{\url{https://github.com/alex88ridolfi/SPIDER\_TWISTER}} and a step size of $d\Tasc = 4.655$~s.
We obtained a clear detection of J1048 in the L-band observation taken in August 2019 when using a $\Tasc$ MJD value of $58721.666409(54)$. This is $\sim28$ s later than the $\Tasc$ value that can be extrapolated from the \cite{Deneva} ephemeris. Folding P band data taken simultaneously with the same parameters, however, resulted in a nondetection. 

The simultaneous SRT+LOFAR observations of J1048 taken in March 2020 were analyzed in the same way. The only difference was that, given the very long duration of these observations (which covered one full orbit), we also splitted them into 1-hr-long chunks and searched for J1048's pulsations separately in each of them.  Still, we were unable to detect J1048 in any of the three radio bands.

\begin{figure}
   \centering
   \resizebox{\hsize}{!}{\includegraphics{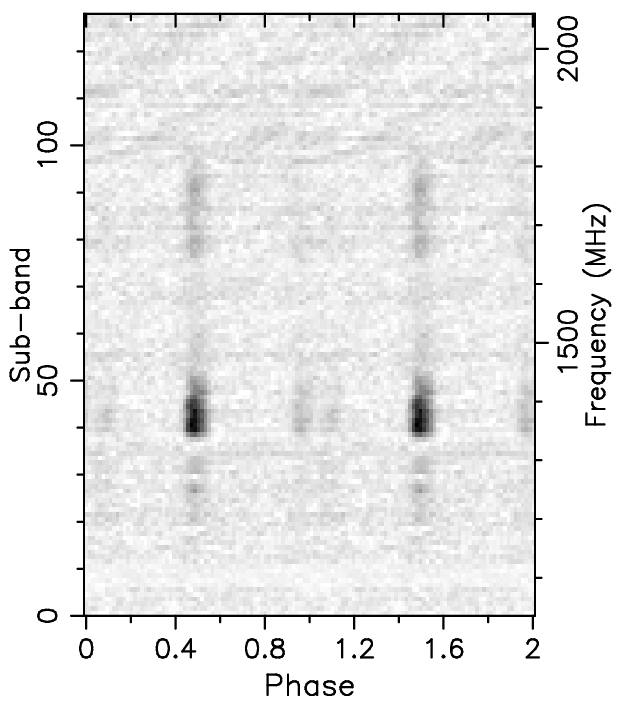}}
      \caption{Intensity as a function of pulse phase (x-axis) and frequency (y-axis) for the SRT observation performed in the L band on 19 August, 2019. Signal strength is not constant at all frequencies. At $\sim$1.4\,GHz there is a peak of intensity, while below 1.3\,GHz and over 1.8\,GHz the pulsed signal is almost absent.}
         \label{Fig_scintillation}
   \end{figure}

The nondetection in the L-band observation of March 2020 can likely be ascribed to interstellar scintillation, which is typically affecting pulsars with very low DMs (such as J1048). This effect randomly enhances or suppresses the observed flux density of the pulsar, with a dependence on the observing frequency. The confirmation that scintillation is indeed important for J1048 is given by the detection of the August 2019 L-band observation (Fig.\,\ref{Fig_scintillation}): we can see that the J1048's pulsed signal is boosted only in a very narrow band around $\sim 1.4$~GHz, whereas it is barely detectable in the rest of the band.
In addition to scintillation, another mechanism responsible for the nondetection in the P and VHF bands, can be the absorption due to the intra-binary eclipsing material. 
Partial or total eclipses seen in many BW and RB systems \citep{Rasio+1989} can be ascribed to the frequency-dependent optical depth of the eclipsing material. The dense layers of ionized material ablated from the donor star are less transparent at low frequencies, determining eclipses in the P and VHF bands \citep{Polzin+2018}. 
As a side effect, at low frequencies, the scattering \citep{Lorimer2004} can smear the individual pulses, making them broader than the pulsar spin period, but not preventing their detection.

From the nondetections, we can calculate an upper limit on the mean flux density, $S^{\rm mean}$, that J1048 must have had at the time of the observations. To do so, we use the modified radiometer equation \citep{Manchester1996, Lorimer2004}
\begin{equation}
S^{\rm mean}=\beta \dfrac{ {\rm (S/N)} \,T_{\rm sys}}{G \sqrt{n_p t_{\rm int} \Delta \nu}}\sqrt{\dfrac{\delta}{1-\delta}}\,,
\end{equation}
where $\beta \gtrsim 1$ is a correction factor due to digitization; $n_p$ is the number of
polarizations summed; S/N is the pulse signal-to-noise ratio; $T_{\rm sys}$ is the system noise temperature; $G$ is the telescope gain; $t_{\rm int}$ is the integration time; $\Delta \nu$ is the bandwidth; $\delta$ is the pulse duty cycle. Table\,\ref{tab:flux density} lists the values of each parameter for all the radio observations that we used in the $S^{\rm mean}$ estimations.  For the August 2019 P band nondetection we estimate $S_{\rm P}^{\rm mean} \lesssim 0.57$\,mJy, whereas in the case of the March 2020 nondetections we find $S_{\rm L}^{\rm mean}  \lesssim 0.05$\,mJy, $S_{\rm P}^{\rm mean}  < 0.24$\,mJy and $S_{\rm VHF}^{\rm mean}  \lesssim 9.21$\,mJy for the P, L and VHF observations, respectively.
For the only radio detection available to us (August 2019 L-band observation) we measure $S_{\rm L}^{\rm mean} \sim 1.22$\,mJy.

 \begin{table*}
 \footnotesize
 \renewcommand{\arraystretch}{1.2}
\centering
\caption{PSR J1048+2339 flux density at 0.1-0.2\,GHz, 0.3-0.4\,GHz and 1.3-1.8\,GHz.}     
\begin{tabular}{c c c c c c c c c}          
\hline
\hline   
\smallskip
Band & $T_{\rm sys}$ & Gain  & $\Delta \nu$& $t_{\rm int}$ & $n_p$ & $\beta$ & $S^{\rm mean}$\\
 & [K] & [K\,Jy$^{-1}$] & [MHz] & [s] & & &  [mJy]  \\  
\hline
L (August 2019) & 30$^{a}$& 0.55$^{a}$& 281& 5400& 2& 1&1.2\\
P (August 2019) & 65$^{a}$&0.52$^{a}$ &60 &5400 &2 &1&$<$0.57 \\
L (March 2020) & 30$^{a}$&0.55$^{a}$ &400 &29960 &2 &1&$<$0.05 \\
P (March 2020) &65$^{a}$ &0.52$^{a}$ & 60&29960 &2 &1 &$<$0.24 \\
0.1-0.2\,GHz (March 2020) & 892$^{b}$ & 0.18$^{c}$& 76 &25200 &2 &1& $<$9.2\\
\hline

\\
\end{tabular}
\label{tab:flux density}

$^{a}$ \url{https://srt-documentation.readthedocs.io/en/latest/antenna.html#lp-band-filters}\\
$^{b}$ \cite{Lawson1987} \\
$^{c}$ \cite{vanHaarlem13}\\
\end{table*}

\section{Optical timing analysis}
\label{optical timing}

We tested the acquired data with IFI+Iqueye for pulsations folding all three observations together with the spin period of the pulsar (0.00466516294153(3)\,s, extrapolated from \citealt{Deneva} at MJD 58927.0) and 10 phase bins. The time of passage at the ascending node was fixed at $\Tasc = 58721.666409$\,MJD (see Sect.\,\ref{radio timing}). A $\chi^2$ test with a constant term corresponding to the average count rate was applied to the folded profile. The $\chi^2$ returns a value of 11.5 for 9 degrees of freedom, corresponding to a false positive probability of 24\% and a significance of pulsed emission of only 1.2$\sigma$. Different choices for the number of phase bins provide similar results.

A conservative upper limit to the optical pulsed emission is derived from the power spectra of the three observations. The maximum excess power in the 0.5-500\,Hz frequency range with a time bin of 1\,ms corresponds to a root mean square fractional variability of $\sim 0.1$\% on the non background subtracted light curve or $\sim 2$\,counts\,s$^{-1}$ for the observed average background count rate. This upper limit refers to a single trial for the time of passage at the ascending node.

We performed also an epoch folding search varying the value of $\Tasc$ for our epoch in an interval between -250\,s and +250\,s around the value reported above and in steps of 1 s, and folding the event lists with 16 bins per period. No significant signal is detected. The maximum of the $\chi^2$ is $\sim 42$ for 15 degrees of freedom, corresponding to a significance of only 1.6$\sigma$ (for 501 trials).
  
\section{Discussion and conclusions}
\label{discussion}

The multiwavelength observations acquired in March 2020 confirmed the extreme emission variability of J1048 detected by \cite{Deneva}, \cite{Cho}, \cite{Strader} and \cite{Yap}. We observed rapid and strong emission changes in the H$\alpha$ emission line along the orbit (Fig.\,\ref{Fig_Ha_variability}). Using the optical photometry we revealed an average $R\sim$19.3\,mag counterpart with orbital variability typical of moderate irradiated systems (Fig.\,\ref{Fig_LC}). The shape of the $R$-band light curve showed that the modulation was dominated by the pulsar heating and not by ellipsoidal variations. We measured with radio timing analysis a strong orbital variation as in other RB systems \citep{Jaodand, Ridolfi}; the $T_{\rm asc}$ lagged the predicted value of \cite{Deneva} ephemeris by $\sim$28\,s.

We measured the radial velocity curve through the cross-correlation with late-type template stars and we set the following constraints to the mass ratio, to the mass of the NS and of the companion star: $0.209 < q < 0.250$, $(1.0 < M_{\rm NS} < 1.6)\,{\rm sin}^{-3}i\,M_{\odot}$ and $(0.24 < M_2 < 0.33)\,{\rm sin}^{-3}i\,M_{\odot}$. Assuming the inclination upper limit of $76^{\circ}$ \citep{Yap}, we derived a NS mass of $1.08 < M_{\rm NS} < 1.76$\,$M_{\odot}$. Given the uncertainty on the inclination, we cannot exclude the presence of a high-mass NS. Assuming the NS mass upper limit of $\sim2.3\,M_{\odot}$, the inclination of J1048 must be $>62^{\circ}$ if the system is nonirradiated, and $>53^{\circ}$ if it is strongly irradiated.
We constrained the spectral type of the donor to a K8\,V star ($\sim$4000\,K). The companion star contributes 56\% to the total light observed at 6400\AA, suggesting the presence of an additional diluting source such as an intra-binary shock.

We presented the first H$\alpha$ Doppler tomography of a RB system in a rotation-powered, radio pulsar state.  Doppler tomograms of RBs were only performed for the two transitional systems PSR\,J1023+0038 and XSS J12270-4859, but during the disk state \citep{Hakala2018, deMartino2014}. The reconstructed H$\alpha$ Doppler map revealed a significant emission close to the inner Lagrangian point along the gravitational free-fall trajectory of the gas and at the companion star surface (see Fig.\,\ref{Fig_dopplermap}). 
The extended structure seen in the H$\alpha$ line is a first and clear evidence of an intra-binary shock emission caused by the interaction of the pulsar wind with material lost from the companion star. Such shocks have been earlier proposed to be responsible for the X-ray orbital variability in MSP binaries \citep{Bogdanov2005} and their presence has been indirectly confirmed by several observations in different domains \citep{Wadiasingh2018}. Our detection of a localized emission of H$\alpha$ line makes J1048 the first RB showing direct evidence of an intra-binary shock emitting
at other wavelengths. Whether the secondary star is overflowing its Roche lobe, the observed emission could be produced by the interaction of the gas streaming the $L_1$ and the pulsar wind. Due to the lack of strong constraints on Roche-lobe filling factor (see \citealt{Yap}), the shock emission could also be due to the interaction of ablated material from the companion and the strong pulsar wind. In this last configuration, and given that the H$\alpha$ emission comes close to $L_1$, this could be a signature of matter from the intra-binary shock that are ducted to the companion surface along the magnetic field of the companion star as hypothesized by \cite{Sanchez2017}. The fact that this system shows variability on different epochs may be compatible with this scenario.

The presence of the intra-binary shock emission can explain the prolonged radio eclipses \citep{Deneva, Deneva2020} and the sporadic appearance of H$\alpha$ double-peaked emission lines \citep{Strader} observed in J1048. This map confirms the hypothesized scenario in which RBs are characterized by a large amount of material being stripped off the companion star due to the pulsar ablation process \citep{Breton2013}. The intra-binary shock in J1048 appears to be closer to the companion than to the pulsar. This evidence contrasts with the recently proposed scenario for the X-ray shock emission in RB systems \citep{Wadiasingh2018, vanderMerwe} in which the intra-binary shock is assumed to be closer to the pulsar. This assumption is supported by the large radio eclipse fractions observed in RBs at pulsar superior conjunction \citep{Archibald, Roy2015, Deneva}, as well as the double-peaked X-ray orbital modulation centered at the pulsar inferior conjunction \citep{Romani2016, AlNoori2018, deMartino2020}. However, the Doppler map of J1048 suggests that X-ray and H$\alpha$ emissions do not trace the same region of the intra-binary shock material. H$\alpha$ emission may trace a portion of intra-binary shock closer to the companion while the
X-rays could originate closer to the NS.


The absence of the pulsed radio signal in March 2020 observations cannot be ascribed to a state transition such as the one observed in the tMSPs \citep{Archibald, Papitto2013}, but rather to scintillation or absorption phenomena \citep{Gedalin1993, Thompson1994, Roy2015}. At very low frequency (0.1-0.2\,GHz) the radio signal could be totally eclipsed by absorption due to the presence of a low-density highly-ionized gas cloud spilling off the companion \citep{Broderick2016}. Indeed, no increase in the X-ray and optical fluxes were observed and the H$\alpha$ Doppler map clearly showed the presence of an intra-binary shock and not an accretion disk. The presence of an extended shock front could lead in the near future to the formation of an accretion disk, but as we have observed in the three tMSPs discovered so far, the state transition is not predictable and takes place in a very rapid (few weeks) timescale  \citep{Archibald, Papitto2013, Bassa2014}.

Although J1048 is one of only a handful of RBs that occasionally displays emission lines in their optical spectra, similar behavior is observed in other RBs and BWs, such as 3FGL\,J0838.8--2829 \citep{Halpern2017a, Halpern2017b}, PSR\,J1628--3205 \citep{Cho, Strader}, and PSR\,J1311--3430 \citep{Romanova2015}. Further spectroscopic observations of these RBs and BWs therefore seem warranted. It would be interesting to monitor these systems for several contiguous orbital phases to study the evolution of the Doppler map and confirm the rapid variability of the gas distribution.

\begin{acknowledgements}
We thank the referee for very useful comments and suggestions.
AMZ acknowledge the support of the PHAROS COST Action (CA16214). AMZ would like to thank G. Benevento for comments on draft and for discussions. We thank Michele Fiori and Giampiero Naletto (part of the Aqueye+Iqueye team) for the support and service activities carried out within the framework of this project. We thank Jorge Casares for providing the stellar templates spectra. SC and PDA acknowledge support from ASI Grant I/004/11/3. AR gratefully acknowledges financial support by the research Grant “iPeska” (P.I. Andrea Possenti) funded under
the INAF national call PRIN-SKA/CTA approved with the Presidential Decree 70/2016. FCZ is supported by a Juan de la Cierva fellowship. AP and FCZ
acknowledge the International Space Science Institute (ISSI-
Beijing), which funded and hosted the international team
`Understanding and Unifying the Gamma-rays Emitting
Scenarios in High Mass and Low Mass X-ray Binaries'. DDM and AP acknowledge financial support from the Italian Space Agency
(ASI) and National Institute for Astrophysics (INAF) under agreements
ASI-INAF I/037/12/0 and ASI-INAF n.2017-14-H.0;  from INAF "Sostegno alla ricerca scientifica main streams dell'INAF", Presidential Decree 43/2018 and  from PRIN-SKA/CTA: Toward the SKA and CTA era" (PI: Giroletti) Presidential Decree 70/2016; and partial support from PHAROS COST Action (CA16214). TMD acknowledges support from the Spanish Ministerio de Ciencia e Innovaci\'on under grant AYA2017-83216-P and the Ram\'on y Cajal Fellowship RYC-2015-18148. LZ acknowledges financial support from the Italian Space Agency (ASI) and National Institute for Astrophysics (INAF) under agreements ASI-INAF I/037/12/0 and ASI-INAF n.2017-14-H.0 and from INAF "Sostegno alla ricerca scientifica main streams dell'INAF" Presidential Decree 43/2018. The Sardinia Radio Telescope is funded by the Department of Universities and Research (MIUR), the Italian Space Agency (ASI), and the Autonomous Region of Sardinia (RAS), and is operated as a National Facility by the National Institute for Astrophysics (INAF). This work made use of data supplied by the UK Swift Science Data Centre at the University of Leicester. The results presented in this paper are based on observations collected at the European Organisation for Astronomical Research in the Southern Hemisphere under ESO programme ID 0104.D-0589(A), on the observations collected at the Galileo telescope (Asiago, Italy) of the University of Padova and on data obtained with the International LOFAR Telescope (ILT) under project code DDT13\_001. LOFAR is the Low
Frequency Array designed and constructed by ASTRON. It has observing, data processing, and data storage facilities in several countries, that are owned by various parties (each with their own funding sources), and that are collectively operated by the ILT foundation under a joint scientific policy. The ILT resources have benefitted from the following recent major funding sources: CNRS-INSU, Observatoire de Paris and Universit\'e d'Orl\'eans, France; BMBF, MIWF-NRW, MPG, Germany; Science Foundation Ireland (SFI), Department
of Business, Enterprise and Innovation (DBEI), Ireland; NWO, The Netherlands; The Science and Technology Facilities Council, UK.   
\end{acknowledgements}

%
%

\bibliography{biblio}

\end{document}